\documentclass[%
 reprint,
 amsmath,amssymb,
 aps,
]{revtex4-2}
\usepackage{dcolumn}
\usepackage{bm}
\usepackage{caption}
\usepackage{subfig}
\usepackage{amsmath}
\usepackage{indentfirst}
\usepackage{float}
\usepackage{graphicx}
\usepackage{hyperref}
\usepackage{mciteplus}
\usepackage{hyperref}
\usepackage{amssymb}
\usepackage{amsthm}
\usepackage{amstext}
\usepackage{epstopdf}
\usepackage{color}
\newcommand\1{{\textbf 1}}
\begin{document}

\preprint{USTC-ICTS/PCFT-20-35}
\title[USTC-ICTS/PCFT-20-35]{\boldmath On Schwinger pair production between D3 branes}

\author{Zihao Wu}%
 \email{wuzihao@mail.ustc.edu.cn}
\affiliation{Interdisciplinary Center for Theoretical Study, University of Science and Technology of China, Hefei, Anhui 230026, China}

\begin{abstract}
We study the open string pair production between two D3 branes, which will give rise to similar effect as Schwinger pair production for observers on one of the D3 branes. The D3 branes are placed parallel at a distance, and they are carrying world-volume electromagnetic fluxes that takes general form. We derive the pair production rate by computing the interaction amplitude between the D3 branes. We discussed how to maximize the pair production rate in this general case. We also mentioned that the general result can be used to describe other system such as D3-D1, where the pair production is ultra large compared to original Schwinger pair production, making it hopeful to observe pair production in experiments.
\end{abstract}
\keywords{string theory, D brane, Schwinger pair production}

\hfill{\small USTC-ICTS/PCFT-20-35}

\maketitle
\section{Introduction}
Schwinger pair production \cite{Schwinger:1951nm} is one of the most impressive effects in Quantum Electrodynamics (QED). It describes electron-positron pairs production in a vacuum stimulated by a strong electromagnetic filed. Although Schwinger pair production has been studied for decades, this effect has not yet been observed in experiments. The reason is that in order to obtain a significant pair production rate, the applied electric field needs to be as strong as $\frac{{m_e}^2c^3}{e \hbar}\approx10^{18}$V/m. This magnitude is far higher than the lab capacity.

\begin{figure}[H]
\centering
\includegraphics[width=0.345\textwidth]{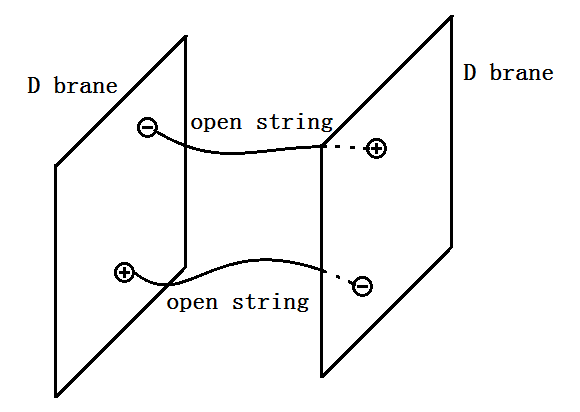}
\caption{Open string pair between D branes}
\label{Open string pair between D branes}
\end{figure}

Although Schwinger pair production in QED is hard to observe in experiments, there is a similar effect arose in string theory:
the open string pair production between D branes, which provides another possibility for observation of pair production.
D branes \cite{Duff:1994an,Schmidhuber:1996fy,DiVecchia:1999fje,DiVecchia:1999mal}, as a kind of soliton in string theory, preserves one half spacetime supersymmetries. When two D branes are placed parallel to each other, without electromagnetic flux on their world-volumes, the net interaction between them vanishes and the system is stable \cite{Polchinski:1995mt}. However, if the D branes carry world-volume fluxes, the system may decay via open string pair production. In such an unstable system, open strings with terminal points ending on the two D branes, separately, will be produced between the D branes (Figure \ref{Open string pair between D branes}). This effect is just like the Schwinger pair production in QED. There were several previous studies on this effect. The bosonic string pair production between D branes carrying electric fluxes is studied in \cite{Burgess:1986dw}. In \cite{Bachas:1992bh}, production of super open strings is studied. {\color{black} In \cite{Acatrinei:2000qm,Porrati:1993qd,Ferrara:1993sq}, magnetic fluxes is taken into consideration. In this paper, we follow the researches in \cite{Lu:2009yx,Lu:2009pe,Lu:2009au,Bolognesi:2012gr,Lu:2017tnm,Lu:2018suj,Lu:2018nsc,Jia:2018mlr,Jia:2019hbr,Lu:2019ynq,Zhang:2020qhk}, studing the open string pair production between D branes, in the system built from type II superstring theory.

As revealed by these previous studies which this paper is based on, one }difference between string pair production and Schwinger pair production is that the pair production rate of open string can be exponentially enhanced by magnetic filed. 
This will provide another possibility for observation as mentioned {\color{black}in the previous studies}. If our (3+1)-dimensional spacetime were considered as a D3 brane, observers on a D3 brane can only see the terminal points of open strings, as charged particles. Open string pair production will look like particle pair production for these observers. The mass of created particles is related to the separation of D branes. If the separation is not too large, and with the help of the magnetic enhancement, it is hopeful for us to observe the pair production of charged particles, in case that the electromagnetic field strength is limited. This may provide a possibility to test the existence of extra dimension as well as verifying the underlying string theory.

\begin{figure}[H]
\centering
\includegraphics[width=0.34\textwidth]{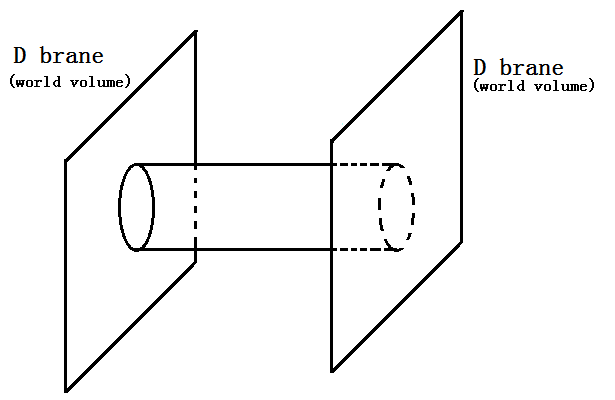}
\caption{The Cylinder Diagram}
\label{The Cylinder Diagram}
\end{figure}

The lowest ordered diagram of the open string pair production is a one-loop cylinder diagram, shown in Figure \ref{The Cylinder Diagram}. This diagram is the same as the tree-diagram where the D branes interacts via exchanging closed strings. In this picture, the interaction amplitude between the D branes can be represented by the electromagnetic fluxes, via the boundary states of them \cite{Billo:1998vr,DiVecchia:1999fje}. By analyzing this amplitude in open string perspective, we can get the expression of open string pair production rate. 
In previous works \cite{Lu:2009yx,Lu:2009au,Lu:2009pe,Lu:2017tnm,Lu:2018suj,Lu:2018nsc,Jia:2018mlr,Jia:2019hbr,Lu:2019ynq,Zhang:2020qhk}, researchers calculated the interaction amplitude, and derived the open string pair production afterwards. In their studies, the electromagnetic fluxes were taken as many different kinds of forms. In this paper, we focus on D3-D3 system.
In such a system, the fluxes have 12 components. We are taking all of them as nonzero values in this paper. The results given in this paper will be expressed in terms of 6 world-volume Lorentz invariants built from the fluxes, making them manifestly world-volume Lorentz invariant. With this result, we will show how to make the pair production rate as large as possible, {\color{black} considering the limited strength of electromagnetic field, to provide more possibility for detection of this phenomenon}

This paper is organized as follows. In section \ref{SecIA} we will derive the general form of interaction amplitude in terms of 6 world-volume Lorentz invariants, and give a brief discussion on the behavior of the interaction amplitude in weak field limit. Then, in section \ref{SecPPR} we give the pair production rate in weak field limit, report the magnetic enhancement and discuss how to make the pair production rate as large as possible. In section \ref{Conclusions} we summarize this paper. Relative detailed calculations and proofs are in the appendix.

\section{The interaction amplitude}\label{SecIA}
In this section we will present the interaction amplitude for the D3-D3 system, which is what we need to derive open string pair production rate in the next section. The closed string cylinder amplitude can be read from \cite{Jia:2019hbr} as
\begin{equation}\label{4-dim IA}
\Gamma=\frac{4\mathcal{F}V_4\sqrt{{\rm det}(1+F){\rm det}(1+F^\prime)}}{(8\pi^2 \alpha^\prime)^2}\int_0^{+\infty}\frac{{\rm d} t}{t^3}e^{-\frac{y^2}{2\pi \alpha^\prime t}}\prod_{n=1}^{+\infty}C_n,
\end{equation}
where $V_4$ is the volume of the world-volumes of D3 branes, $F$ and $F^\prime$ are the respective world-volume fluxes on the two D3 branes, which are anti-symmetric, $y$ is the separation between two D branes and $|z|=e^{-\pi t}$. In the above, $\mathcal{F}=({\rm cos}\pi \nu_0-{\rm cos}\pi \nu_1)^2$, and
\begin{widetext}
\begin{equation}\label{21Cn}
\begin{aligned}
C_n=&\frac{(1-2 |z|^{2n}{\rm cos}\pi(\nu_0+\nu_1)+|z|^{4n})^2(1-2|z|^{2n}{\rm cos}\pi(\nu_0-\nu_1)+|z|^{4n})^2}{(1-|z|^{2n})^4(1-2|z|^{2n}{\rm cos}2\pi \nu_0+|z|^{4n})(1-2|z|^{2n}{\rm cos}2\pi \nu_1+|z|^{4n})}.
\end{aligned}
\end{equation}
\end{widetext}
The two parameters $\nu_0$ and $\nu_1$ are determined in terms of $F$ and $F^\prime$ via the eigenvalues of a certain matrix $W$, which are $\lambda_0$, ${\lambda_0}^{-1}$, $\lambda_1$ and ${\lambda_1}^{-1}$, with $\lambda_0=e^{2\pi i \nu_0}$ and $\lambda_1=e^{2\pi i \nu_1}$, where $W$ is defined as

\begin{equation}\label{1W}
W=M{M^\prime}^{\rm T},
\end{equation}
with
\begin{equation}\label{2M}
M=(\1-F)(\1+F)^{-1},
\end{equation}
and similar for $M^\prime$ with $F$ replaced by $F^\prime$. Here, transpose of a matrix is performed under the flat Minkowski metric, which means that 
\begin{equation}
{(M^{\rm T})_\mu}^\nu={M^\nu}_\mu=\eta_{\mu\sigma}{M_\rho}^\sigma \eta^{\rho\nu},
\end{equation}
where $\eta_{\mu\nu}$ is the Minkowski metric. Additionally, the default index configuration of a matrix is like ${M_\mu}^\nu$ in this paper. Note that $W$, $M$ and $M^\prime$ are special orthogonal matrices, i.e., ${W_\mu}^\rho {(W^{\rm T})_\rho}^\nu={\delta_\mu}^\nu$, ${\rm det}W=1$ and similar for $M$ and $M^\prime$. As given in the appendix of \cite{Jia:2019hbr}, $W$ can always be diagonalized with four eigenvalues mentioned before, which are $\lambda_0$, ${\lambda_0}^{-1}$, $\lambda_1$ and ${\lambda_1}^{-1}$, with $\lambda_0=e^{2\pi i \nu_0}$ and $\lambda_1=e^{2\pi i \nu_1}$. In addition, among the parameters $\nu_0$ and $\nu_1$, one of them is real, and the other one is either pure imaginary or zero. In Appendix \ref{Appendix B} of this paper, we will provide a different proof of this. Note that the amplitude \eqref{4-dim IA} is invariant under $\nu_0 \leftrightarrow \nu_1$ and $\nu_\alpha \to \nu_\alpha+1$, where $\alpha=$0 or 1, we can set $\nu_0=i\bar{\nu}_0$, with $0\leq\bar{\nu}_0<\infty$ and $0\leq\nu_1\leq1$ as our conventions in this paper. As a consequence, we have $\mathcal{F}=({\rm cosh}\pi {\bar{\nu}_0}-{\rm cos}\pi \nu_1)^2\geq0$.

In order to express the interaction amplitude \eqref{4-dim IA} in terms of the fluxes, we need to express the eigenvalues of $W$, or equivalently, $\nu_0$ and $\nu_1$ in terms of the fluxes. In practice, we will instead calculate the combinations of these eigenvalues as
\begin{small} 
\begin{equation}\label{52c2c2}
\begin{aligned}
{\rm cos}^2{\pi}\nu_0{\rm cos}^2{\pi}\nu_1=&\frac{1}{4}({\rm cos}2 \pi \nu_0+1)({\rm cos}2 \pi \nu_1+1)\\
=&\frac{1}{16}(\lambda_0+{\lambda_0}^{-1}+2)(\lambda_1+{\lambda_1}^{-1}+2)\\
=&\frac{1}{16}\Big(\frac{1}{2}({\rm Tr}W)^2-\frac{1}{2}{\rm Tr}(W^2)+2{\rm Tr}W+2\Big),
\end{aligned}
\end{equation}
\end{small} 
and
\begin{small} 
\begin{equation}\label{53s2s2}
\begin{aligned}
{\rm sin}^2{\pi}\nu_0{\rm sin}^2{\pi}\nu_1=&\frac{1}{4}({\rm cos}2 \pi \nu_0-1)({\rm cos}2 \pi \nu_1-1)\\
=&\frac{1}{16}(\lambda_0+{\lambda_0}^{-1}-2)(\lambda_1+{\lambda_1}^{-1}-2)\\
=&\frac{1}{16}\Big(\frac{1}{2}({\rm Tr}W)^2-\frac{1}{2}{\rm Tr}(W^2)-2{\rm Tr}W+2\Big).
\end{aligned}
\end{equation}
\end{small} 
Therefore, we only need to calculate ${\rm Tr}W$ and $({\rm Tr}W)^2-{\rm Tr}(W^2)$. Since these results are invariant under Lorentz transformations, we can express them in terms of several Lorentz invariants built from the fluxes $F$ and $F^\prime$. In what follows, we will define these Lorentz invariants and express the left-hand sides of \eqref{52c2c2} and \eqref{53s2s2} in terms of them. With these, we will express the interaction amplitude in terms of the Lorentz invariants and discuss its related properties.
\subsection{The interaction amplitude in terms of Lorentz invariants}
In general, the world-volume flux on each of the D3 branes can be expressed as
\begin{equation}\label{15F}
F=
\begin{pmatrix}
0&f_1&f_2&f_3\\
f_1&0&-g_3&g_2\\
f_2&g_3&0&-g_1\\
f_3&-g_2&g_1&0
\end{pmatrix},
\end{equation}
similar for $F^\prime$, with $f_i$ and $g_i$ replaced by ${f_i}^\prime$ and ${g_i}^\prime$, where $i=1,2,3$. We introduce the Hodge dual of $F$ as
\begin{equation}
(\star F)^{\alpha\beta}=\frac{1}{2}\epsilon^{\alpha\beta\gamma\delta}F_{\gamma\delta}.
\end{equation} 
That is
\begin{equation}\label{17*F}
{(\star F)_\alpha}^\beta=
\begin{pmatrix}
0&g_1&g_2&g_3\\
g_1&0&f_3&-f_2\\
g_2&-f_3&0&f_1\\
g_3&f_2&-f_1&0
\end{pmatrix}.
\end{equation}
The definition of $\star F^\prime$ is similar to $\star F$. With these, we define the Lorentz invariants as

\begin{equation}\label{LiTr}
\begin{aligned}
&a=\frac{1}{2}{\rm Tr}(F^2),\\
&a^\prime=\frac{1}{2}{\rm Tr}({F^\prime}^2),\\
&b=\frac{1}{4}{\rm Tr}(F(\star F)),\\
&b^\prime=\frac{1}{4}{\rm Tr}({F^\prime}(\star {F^\prime})),\\
&c=\frac{1}{2}{\rm Tr}({F^\prime}F),\\
&d=\frac{1}{2}{\rm Tr}(F(\star {F^\prime})).\\
\end{aligned}
\end{equation}
Since the right-hand sides of \eqref{LiTr} are manifestly Lorentz invariant, the six variables we defined in the left-hand sides of \eqref{LiTr} are indeed invariants under Lorentz transformations. These Lorentz invariants can also be expressed in terms of components of $F$ and $F^\prime$ as
\begin{equation}\label{68Li}
\begin{aligned}
&a=(\vec{f})^2-(\vec{g})^2,\\
&a^\prime=(\vec{f^\prime})^2-(\vec{g^\prime})^2,\\
&b=\vec{f}\cdot\vec{g},\\
&b^\prime=\vec{f^\prime}\cdot\vec{g^\prime},\\
&c=\vec{f}\cdot\vec{f^\prime}-\vec{g}\cdot\vec{g^\prime},\\
&d=\vec{f}\cdot\vec{g^\prime}+\vec{f^\prime}\cdot\vec{g},
\end{aligned}
\end{equation}
where $\vec{f}\cdot\vec{g}$ means $f^ig_i$ where $i$ sum over from 1 to 3, $(\vec{f})^2$ means $\vec{f}\cdot\vec{f}$, similar for other terms. We can express the left-hand sides of \eqref{52c2c2} and \eqref{53s2s2} in terms of the six Lorentz invariants as
\begin{equation}\label{40ccss}
\begin{aligned}
&{\rm cos}^2\pi \nu_0{\rm cos}^2\pi \nu_1=QQ^\prime(1-c-bb^\prime)^2,\\
&{\rm sin}^2\pi \nu_0{\rm sin}^2\pi \nu_1=-QQ^\prime(\Delta b)^2,
\end{aligned}
\end{equation}
where 
\begin{equation}
\Delta b\overset{\rm def}{=}b+b^\prime-d=\Delta \vec{f}\cdot\Delta \vec{g},
\end{equation}
where $\Delta \vec{f}=\vec{f}-\vec{f^\prime}$ and $\Delta \vec{g}=\vec{g}-\vec{g^\prime}$, and
\begin{equation}\label{69 Q}
Q\overset{\rm def}{=}\frac{1}{{\rm det}(\1+F)}=\frac{1}{1-a-b^2},
\end{equation}
similar for definition of $Q^\prime$ with a prime added on each variable. The detailed derivation of \eqref{40ccss} is given in Appendix \ref{Appendix A}. Note that ${\rm sin}^2\pi \nu_0{\rm sin}^2\pi \nu_1\leq0$ is consistent with what we have claimed that $\nu_0$ and $\nu_1$ are one pure imaginary or zero, with the other one being real. Since in our convention we have chosen $\nu_0=i{\bar{\nu}_0}$, with ${\bar{\nu}_0}$ and $\nu_1$ being real, we can rewrite \eqref{40ccss} as
\begin{equation}
\begin{aligned}
&{\rm cosh}^2\pi {\bar{\nu}_0}{\rm cos}^2\pi \nu_1=QQ^\prime(1-c-bb^\prime)^2,\\
&{\rm sinh}^2\pi {\bar{\nu}_0}{\rm sin}^2\pi \nu_1=QQ^\prime(\Delta b)^2.
\end{aligned}
\end{equation}
We will need the square root of these results. Remember that in our convention, we have ${\bar{\nu}_0}>0$ and $0\leq \nu_1\leq1$ and when there is no electromagnetic flux, we require ${\bar{\nu}_0}=\nu_1=0$. In such a convention, together with \eqref{69 Q}, we have
\begin{equation}\label{90cc}
{\rm cosh}{\pi}{\bar{\nu}_0}{\rm cos}{\pi}\nu_1=\frac{1-c-bb^\prime}{\sqrt{(1-a-b^2)(1-a^\prime-{b^\prime}^2)}},
\end{equation}
and
\begin{equation}\label{91ss}
{\rm sinh}{\pi}{\bar{\nu}_0}{\rm sin}{\pi}\nu_1=\frac{|\Delta b|}{\sqrt{(1-a-b^2)(1-a^\prime-{b^\prime}^2)}}.
\end{equation}

Substituting \eqref{90cc} and \eqref{91ss} into \eqref{4-dim IA}, we get interaction amplitude as
\begin{equation}\label{98IA}
\begin{aligned}
\Gamma&=\frac{4\mathcal{F}V_{4}\sqrt{(1-a-b^2)(1-a^\prime-{b^\prime}^2)}}{(8\pi^2 \alpha^\prime)^2}\\
&\times\int_0^{+\infty}\frac{{\rm d} t}{t^3}e^{-\frac{y^2}{2\pi \alpha^\prime t}}\prod_{n=1}^{+\infty}C_n,
\end{aligned}
\end{equation}
where $C_n$ is given in \eqref{21Cn}. According to \cite{Lu:2018suj}, if the brane separation $y$ is very large compared to string scale, the factor $e^{-\frac{y^2}{2\pi \alpha^\prime t}}$ in the integrand in \eqref{98IA} makes the small-$t$ integration unimportant. Therefore, for large $t$, we have $C_n\approx1$ and
\begin{equation}\label{101V}
\Gamma\approx\frac{V_{4}({\rm cosh}\pi {\bar{\nu}_0}-{\rm cos}\pi \nu_1)^2\sqrt{(1-a-b^2)(1-a^\prime-{b^\prime}^2)}}{4\pi^2y^4}\geq0.
\end{equation}
Under such a condition, we have $\Gamma=0$ if and only if ${\bar{\nu}_0}=\nu_1=0$, which means that when there is no flux on the D3 branes, the net interaction indeed vanishes. This agrees with our former claim. Another example is that when the fluxes on two D3 branes are identical, that is, $F=F^\prime$, we have ${\bar{\nu}_0}=\nu_1=0$ and there will be no net interaction between the D3 branes. If ${\bar{\nu}_0},\nu_1\neq0$, the fact $\Gamma>0$ indicates that the interaction is in general attractive between two D3 branes. Moreover, $\Gamma\sim y^{-4}$ is also expected, since there are in total 6 Dirichlet-Dirichlet directions, which are perpendicular to the D3 branes. This is an analog to Gauss's theorem which describes the interaction between two charged particles in 3-dimensional space. These results are consistent with \cite{Lu:2018suj}.

\subsection{Weak field limit}\label{secwfl}
It is necessary to consider our results in weak field limit where $F_{\mu\nu}\ll1$ and $F^\prime_{\mu\nu}\ll1$, since the electromagnetic fluxes are generally far less than string scale. In our convention, ${\bar{\nu}_0}$ and $\nu_1$ must be small in such a limit. We will further express ${\bar{\nu}_0}$ and $\nu_1$ directly in terms of the fluxes. Then, we will give the results of interaction amplitude in such a limit.

If we take the lowest order, \eqref{90cc} and \eqref{91ss} becomes
\begin{equation}
\pi^2({{\bar{\nu}_0}}^2-{\nu_1}^2)=a+a^\prime-2c=|\Delta\vec{f}|^2-|\Delta\vec{g}|^2,
\end{equation}
and
\begin{equation}
\pi^2{\bar{\nu}_0}\nu_1=|\Delta b|=|\Delta\vec{f}\cdot\Delta\vec{g}|=|\Delta\vec{f}||\Delta\vec{g}||{\rm cos}\theta|,
\end{equation}
where $\theta$ is the angle between $\Delta\vec{f}$ and $\Delta\vec{g}$. Solving these equations, we get
\begin{widetext}
\begin{equation}\label{54u0}
{\bar{\nu}_0}=\frac{1}{\sqrt{2}\pi}\sqrt{|\Delta\vec{f}|^2-|\Delta\vec{g}|^2+\sqrt{|\Delta\vec{f}|^4+|\Delta\vec{g}|^4+2|\Delta\vec{f}|^2|\Delta\vec{g}|^2{\rm cos}2\theta}},
\end{equation}
and
\begin{equation}\label{55u1}
\nu_1=\frac{1}{\sqrt{2}\pi}\sqrt{|\Delta\vec{g}|^2-|\Delta\vec{f}|^2+\sqrt{|\Delta\vec{f}|^4+|\Delta\vec{g}|^4+2|\Delta\vec{f}|^2|\Delta\vec{g}|^2{\rm cos}2\theta}}.
\end{equation}
\end{widetext}
Substituting \eqref{54u0} and \eqref{55u1} into \eqref{98IA} and keeping the lowest order, we get the interaction amplitude in weak field limit as
\begin{equation}\label{52IAinWFL}
\begin{aligned}
\Gamma\approx&\frac{V_{4}\Big(|\Delta\vec{f}|^4+|\Delta\vec{g}|^4+2|\Delta\vec{f}|^2|\Delta\vec{g}|^2{\rm cos}2\theta\Big)}{16\pi^2y^4}\\,
\end{aligned}
\end{equation}
where we have used $C_n\approx1$ in weak field limit. In this case, the interaction vanishes if and only if $|\Delta\vec{f}|=|\Delta\vec{g}|$ and $\Delta\vec{f}\bot\Delta\vec{g}$. Additionally, we found that \eqref{52IAinWFL} can be written as
\begin{equation}
\begin{aligned}
\Gamma=\frac{V_{4}\Big[\big(\frac{1}{2}(|\Delta\vec{f}|^2+|\Delta\vec{g}|^2)\big)^2-(|\Delta\vec{f}||\Delta\vec{g}|{\rm sin}\theta)^2\Big]}{4\pi^2y^4}\\,
\end{aligned}
\end{equation}
whose numerator is somehow related to the energy density and momentum density of the electromagnetic fluxes.\\[100pt]
\section{The pair production rate}\label{SecPPR}
The open string pair production rate can be derived from the result of interaction amplitude acquired in the last section. The way of doing this is already introduced in \cite{Jia:2019hbr}. So, here we just refer to it and express the open string pair production rate $\mathcal{W}$ as
\begin{widetext}
\begin{equation}
\begin{aligned}
\mathcal{W}=&\frac{8{\rm sinh}{\pi}{\bar{\nu}_0}{\rm sin}{\pi}\nu_1\sqrt{{\rm det}(1+F){\rm det}(1+F^\prime)}}{(8\pi^2 \alpha^\prime)^{2}}e^{-\frac{y^2}{2\pi \alpha^\prime {\bar{\nu}_0}}}\frac{(1+{\rm cosh}\pi \frac{\nu_1}{{\bar{\nu}_0}})^2}{{\rm sinh}\pi \frac{\nu_1}{{\bar{\nu}_0}}}\prod_{n=1}^\infty Z_n(\frac{1}{{\bar{\nu}_0}}),
\end{aligned}
\end{equation}
where
\begin{equation}
Z_n(t)=\frac{(1-2|z|^{2n}{\rm cosh}\pi (\nu_0+\nu_1)t+|z|^{4n})^2(1-2|z|^{2n}{\rm cosh}\pi (\nu_0-\nu_1)t+|z|^{4n})^2}{(1-|z|^{2n})^4(1-2|z|^{2n}{\rm cosh}\pi 2\nu_0t+|z|^{4n})(1-2|z|^{2n}{\rm cosh}\pi 2\nu_1t+|z|^{4n})},
\end{equation}
where $|z|=e^{-\pi t}$.
\end{widetext}
We need to express the parameters $\bar{\nu}_0$, $\nu_1$ and the determinants in terms of fluxes. We firstly use \eqref{91ss}, and get
\begin{equation}\label{99PPR}
\begin{aligned}
\mathcal{W}=&\frac{8|\Delta b|}{(8\pi^2 \alpha^\prime)^{2}}e^{-\frac{y^2}{2\pi \alpha^\prime {\bar{\nu}_0}}}\frac{(1+{\rm cosh}\pi \frac{\nu_1}{{\bar{\nu}_0}})^2}{{\rm sinh}\pi \frac{\nu_1}{{\bar{\nu}_0}}}\prod_{n=1}^\infty Z_n(\frac{1}{{\bar{\nu}_0}}).
\end{aligned}
\end{equation}
Next, we need to consider two issues. One issue is that, considering the reality, the electromagnetic field we can apply in experiments is far less than string scale. Thus, the weak field limit introduced in section \ref{secwfl} should be taken into consideration. In such a limit, we have $\bar{\nu}_0,\nu_1\to0$, so $Z_n(\frac{1}{{\bar{\nu}_0}})\to1$. Thus,
\begin{equation}\label{4-dim PPR in WFL}
\mathcal{W}=\frac{{\bar{\nu}_0}\nu_1}{8\pi^2 {\alpha^\prime}^{2}}e^{-\frac{y^2}{2\pi \alpha^\prime {\bar{\nu}_0}}}\frac{\big(1+{\rm cosh}\pi \frac{\nu_1}{{\bar{\nu}_0}}\big)^2}{{\rm sinh}\pi \frac{\nu_1}{{\bar{\nu}_0}}}.
\end{equation}

The other issue is that we hope to create an as-large-as-possible pair production rate to make it easier to observe pair production in experiments. According to \eqref{4-dim PPR in WFL}, ${\bar{\nu}_0}$ and $\nu_1$ should be as large as possible, if the tachyon-free condition (see \cite{Banks:1995ch,Pesando:1999hm,Sen:1999xm,Lu:2007kv,Lu:2018suj}) $y>\sqrt{2\pi^2\alpha^\prime\nu_1}$ is satisfied. Referring our results in section \ref{secwfl}, there are 3 variables, $|\Delta\vec{f}|$, $|\Delta\vec{g}|$ and $\theta$, controlling ${\bar{\nu}_0}$ and $\nu_1$ through \eqref{54u0} and \eqref{55u1}. If $|\Delta\vec{f}|$ and $|\Delta\vec{g}|$ are fixed, the larger ${\rm cos}2\theta$ is, the larger both ${\bar{\nu}_0}$ and $\nu_1$ are. Thus, we prefer to make ${\rm cos}2\theta=1$, i.e. $\theta=0$ or $\theta=\pi$. Then, we get
\begin{equation}\label{109f}
{\bar{\nu}_0}=\frac{1}{\pi}|\Delta\vec{f}|,
\end{equation}
and
\begin{equation}\label{110g}
\nu_1=\frac{1}{\pi}|\Delta\vec{g}|.
\end{equation}
Additionally, if originally $\Delta\vec{f}$ and $\Delta\vec{g}$ does not take the same or opposite directions, we can always find a Lorentz boost to set them so. But this will simultaneously change values of $|\Delta\vec{f}|$ and $|\Delta\vec{g}|$. According to this, we can analyze the pair production rate in case of $\theta=0$ or $\theta=\pi$ in the following discussion.

Substituting \eqref{109f} and \eqref{110g} to \eqref{4-dim PPR in WFL}, we have
\begin{equation}\label{4-dim PPR in WFL by fg}
\mathcal{W}=\frac{|\Delta\vec{f}||\Delta\vec{g}|}{8\pi^4 {\alpha^\prime}^{2}}e^{-\frac{y^2}{2\alpha^\prime |\Delta\vec{f}|}}\frac{\big(1+{\rm cosh}\pi \frac{|\Delta\vec{g}|}{|\Delta\vec{f}|}\big)^2}{{\rm sinh}\pi \frac{|\Delta\vec{g}|}{|\Delta\vec{f}|}}.
\end{equation}
This result agrees with the case where the electric and magnetic fields chosen to be parallel to each other in \cite{Lu:2018nsc}. In this paper, after deriving the result in Lorentz invariant form, we proved that this is the best choice of fluxes, in sense of obtaining the maximum of pair production rate, {\color{black}considering the limited field strengths. As a consequence, the resulting field theory of this limit in the general case also agrees those discussed in \cite{Lu:2018nsc}.}

After proving that choosing electric and magnetic field is our best choice, we continue to discuss how much a pair production can be created by this mechanism. Since the tachyon-free condition now becomes $y>\sqrt{2\pi\alpha^\prime|\Delta\vec{g}|}$, from \eqref{4-dim PPR in WFL by fg} we can learn that for a fixed $|\Delta\vec{g}|$, $\mathcal{W}$ increases if $|\Delta\vec{f}|$ increases, and for a fixed $|\Delta\vec{f}|$, $\mathcal{W}$ increases if $|\Delta\vec{g}|$ increases. {\color{black}This suggests that our field strength should be taken as large as possible}, trying to create an as-large-as-possible pair production rate. 

If we have taken the best condition: electric field and magnetic field been parallel and maximized, is it enough for observation? At least, this open string pair production is much more hopeful to be observed than original Schwinger mechanism, because there exists a magnetic enhancement in \eqref{4-dim PPR in WFL by fg}, also reported in \cite{Lu:2018nsc,Lu:2018suj,Lu:2017tnm,Lu:2009au}. Consider that if we can apply an ultra large magnetic field, such that $\frac{|\Delta\vec{g}|}{|\Delta\vec{f}|}\gg1$, we will have
\begin{equation}\label{ehcmt}
\begin{aligned}
\frac{\mathcal{W}(|\Delta\vec{g}|\neq0)}{\mathcal{W}(|\Delta\vec{g}|=0)}\approx\frac{\pi}{8}\frac{|\Delta\vec{g}|}{|\Delta\vec{f}|}e^{\pi\frac{|\Delta\vec{g}|}{|\Delta\vec{f}|}},
\end{aligned}
\end{equation}
which is what we mentioned before: the open string pair production can be exponentially enhanced by magnetic field, being the main difference between the original Schwinger pair production in QED. Suppose the magnetic field is larger than the electric field by one degree of magnitude, which is $\frac{|\Delta\vec{g}|}{|\Delta\vec{f}|}\sim10$, we result in an enhancement $\frac{\pi}{8}\frac{|\Delta\vec{g}|}{|\Delta\vec{f}|}e^{\pi\frac{|\Delta\vec{g}|}{|\Delta\vec{f}|}}\sim10^{14.2}$. This enhancement will decrease the critical electric field we need to apply to create a detectable pair production. {\color{black}Usually, $|\Delta\vec{f}|$ is small, then $|\Delta\vec{g}|$ should be very large. Although it is also very hard to apply an ultra large magnetic field,} \cite{Lu:2019ynq} gave a discussion on D3-D1 system, where the D1 branes can be considered to be a D3 brane carrying a magnetic flux that $g^\prime\to\infty$. Taken the fluxes other than $g^\prime$ are small, we have $\bar{\nu}_0\to0$ and $\nu_1\to\frac{1}{2}$. Then, \eqref{4-dim PPR in WFL} holds and since $\frac{\nu_1}{\bar{\nu}_0}\gg 1$,  such a system will have an ultra large magnetic enhancement, making it possible for observations of pair production.

\section{Summary}\label{Conclusions}
In this paper we report the open string pair production, as an analogue of Schwinger pair production in QED. It is a substitution mechanism describing particle-anti particle pair production in the vacuum, at presence of electromagnetic field. If such a mechanism exists, experimental detection of pair production is more hopeful, since its production rate is enhanced by the presence of magnetic field. 

We derive the result of interaction amplitude and pair production rate, considering the case where the electromagnetic fluxes take the most general form of D3-D3 system. The interaction amplitude obeys Gauss's law and vanishes if and only if $|\Delta\vec{f}|=|\Delta\vec{g}|$ and $\Delta\vec{f}\bot\Delta\vec{g}$. 

The open string pair production rate is expressed in terms of fluxes in \eqref{4-dim PPR in WFL}. In our discussion we proved that taking $\Delta\vec{f}$ and $\Delta\vec{g}$ in same or opposite direction is the best choice to ensure a maximized pair production rate, if $|\Delta\vec{f}|$ and $|\Delta\vec{g}|$ are already maximized. Afterward, the same exponential magnetic enhancement as discussed in previous works is reported.

The general result of pair production rate in D3-D3 system can also be used to describe D3-D1 system, because the D1 branes behaves like a D3 brane carrying an infinity large magnetic flux. In that system, the magnetic enhancement is large enough. If we are in such a system, we will have higher possibility to produce particle-anti particle pairs in experiments. This will be a way to detect extra dimensions and to verify the underlying string theory.
\acknowledgments
The author would like to thank J.X. Lu, Qiang Jia and Xiaoying Zhu for important discussions. The author acknowledges support by grants from the NSF of China with Grant No:
11235010 and 11775212.
\appendix
\section{Derivation of eigenvalues}\label{Appendix A}
\indent In this appendix we will give a detailed derivation of \eqref{40ccss}. We firstly calculate Tr$W$. From \eqref{1W} and \eqref{2M}, the trace of $W$ can be directly expressed in terms of traces of $F$ and $F^\prime$ as
\begin{equation}\label{terms of TrW}
\begin{aligned}
{\rm Tr}W=&4{\rm Tr}(\1+F)^{-1}(\1-F^\prime)^{-1}-2{\rm Tr}(\1+F)^{-1}\\&-2{\rm Tr}(\1-F^\prime)^{-1}+4,
\end{aligned}
\end{equation}
where we have used $(\1-F)(\1+F)^{-1}=2(\1+F)^{-1}-\1$ and $F^{\rm T}=-F$ and similar for $F^\prime$.
From \eqref{15F}, \eqref{17*F} and \eqref{68Li}, one can check that the world-volume fluxes and their Hodge dual  satisfy the following relations.
\begin{equation}\label{61F*F}
\begin{aligned}
&F^2-(\star F)^2=a\1,\\
&{F^\prime}^2-(\star F^\prime)^2=a^\prime\1,\\
&(\star F)F=F(\star F)=b\1,\\
&(\star F^\prime)F^\prime=F^\prime(\star F^\prime)=b^\prime\1,\\
&FF^\prime-(\star F^\prime)(\star F)=F^\prime F-(\star F)(\star F^\prime)=c\1,\\
&(\star F^\prime)F+(\star F)F^\prime=F^\prime(\star F)+F(\star F^\prime)=d\1,
\end{aligned}
\end{equation}
where $\1$ is the $4\times 4$ identity matrix.
From \eqref{61F*F}, we can derive
\begin{equation}\label{recursion of F}
F^4=F(\star F)^2F+aF^2=aF^2+b^2\1,
\end{equation}
and for $F^\prime$ we have a similar relation with a prime added on each variable. We can calculate the first term in \eqref{terms of TrW} by the power expansion,
\begin{equation}\label{36STmn}
{\rm Tr}(\1+F)^{-1}(\1-F^\prime)^{-1}=\sum_{m,n=0}^{+\infty}(-1)^mT_{m,n},
\end{equation}
where $T_{m,n}$ stands for ${\rm Tr}(F^m{F^\prime}^n)$. Using \eqref{recursion of F}, we have recursion relations as
\begin{equation}\label{65TT}
\begin{aligned}
&T_{m,n}=aT_{m-2,n}+b^2T_{m-4,n},\\
&T_{m,n}=a^\prime T_{m,n-2}+{b^\prime}^2T_{m,n-4}.
\end{aligned}
\end{equation}
The first few terms of this recursion can be directly computed from \eqref{15F} and \eqref{17*F} as
\begin{equation}\label{66T0}
\begin{aligned}
&T_{0,0}=4, \quad T_{2,0}=2a, \quad T_{0,2}=2a^\prime, \quad T_{2,2}=aa^\prime+c^2+d^2,\\
&T_{-1,-1}=-\frac{2c}{bb^\prime},\quad T_{-1,1}=\frac{2d}{b},\quad T_{1,-1}=\frac{2d}{b^\prime},\quad T_{1,1}=2c,
\end{aligned}
\end{equation}
where we have used $\star F=bF^{-1}$ and $\star F^\prime=b^\prime {F^\prime}^{-1}$. Here we choose $m,n$ to start from ($-1$) because these terms are easier to compute. Notice that although here we require $b,b^\prime\neq0$, our final result still holds even if $b$ or $b^\prime$ is 0. From \eqref{65TT} and \eqref{66T0}, deriving $T_{mn}$ is just a typical second-order recursion problem. The results are

\begin{equation}\label{67 Tmn}
\begin{aligned}
&T_{2i,2j}=(aa^\prime+c^2+d^2)Q_i{Q^\prime}_j+2a^\prime b^2Q_{i-1}{Q^\prime}_j\\&\quad\quad\quad\quad+2a{b^\prime}^2Q_i{Q^\prime}_{j-1}+4b^2{b^\prime}^2Q_{i-1}{Q^\prime}_{j-1},\\
&T_{2i-1,2j-1}=2cQ_i{Q^\prime}_j+2dbQ_{i-1}{Q^\prime}_j\\&\quad\quad\quad\quad\quad+2d{b^\prime}Q_i{Q^\prime}_{j-1}-2cbb^\prime Q_{i-1}{Q^\prime}_{j-1},
\end{aligned}
\end{equation}
and, since $F$ and $F^\prime$ are antisymmetric, $T_{2i,2j-1}=T_{2i-1,2j}=0$, where $i$ and $j$ are integers. In the above formula, we have defined
\begin{equation}\label{26Qi}
Q_i=\frac{\alpha^i-\beta^i}{\alpha-\beta},
\end{equation}
where $\alpha^i$ means $\alpha$ to the power of $i$, similar for $\beta^i$. $\alpha$ and $\beta$ are defined as $\alpha+\beta=a$, $\alpha\beta=-b^2$. The definition of ${Q^\prime}_i$ is similar. We sum over $i$ in \eqref{26Qi} and get
\begin{equation}
\sum_{i=0}^{+\infty}Q_i=\frac{1}{1-a-b^2}=Q.
\end{equation}
Similarly, we have $\sum_{i=0}^{+\infty}{Q_i}^\prime=Q^\prime$. From \eqref{36STmn} and \eqref{67 Tmn} we have
\begin{equation}\label{equationNo70}
\begin{aligned}
&{\rm Tr}(\1+F)^{-1}(\1-F^\prime)^{-1}=\sum_{i,j=0}^{+\infty}T_{2i,2j}-\sum_{i,j=0}^{+\infty}T_{2i-1,2j-1}\\
&=QQ^\prime\big[(a-2)(a^\prime-2)+c^2+d^2-2c(1-bb^\prime)-d(b+b^\prime)\big].
\end{aligned}
\end{equation}

Using the same method, we can compute the other terms in \eqref{terms of TrW},
\begin{equation}\label{equationNo71}
\begin{aligned}
&{\rm Tr}(\1+F)^{-1}=\sum_{i=0}^{+\infty}T_{2i,0}=2Q(2-a),\\
&{\rm Tr}(\1-F^\prime)^{-1}=\sum_{i=0}^{+\infty}T_{0,2i}=2Q^\prime(2-a^\prime).
\end{aligned}
\end{equation}

Substituting \eqref{equationNo70} and \eqref{equationNo71} back to \eqref{terms of TrW}, we can derive the result of Tr$W$ as

\begin{equation}\label{72 TrW}
\begin{aligned}
{\rm Tr}W=&4QQ^\prime\big[(2-a)(2-a^\prime)+c^2+d^2-2c(1-bb^\prime)\\
&-2d(b+b^\prime)\big]-4Q(2-a)-4Q^\prime(2-a^\prime)+4\\
=&4QQ^\prime\big[(b^2+1)({b^\prime}^2+1)+c^2+d^2-2c(1-bb^\prime)\\&-d(b+b^\prime)\big].
\end{aligned}
\end{equation}

When deriving the expression after the second equal sign of \eqref{72 TrW}, we have used $(1-a-b^2)Q=(1-a^\prime-{b^\prime}^2)Q^\prime=1$.

\indent Using the same technique, ${\rm Tr}(W^2)$ can also be expressed in terms of traces of $F$ and $F^\prime$ as

\begin{equation}\label{73 TrW2 expanded}
\begin{aligned}
{\rm Tr}(W^2)=&16{\rm Tr}(\1+F)^{-1}(\1-F^\prime)^{-1}(\1+F)^{-1}(\1-F^\prime)^{-1}\\
&-16{\rm Tr}(\1+F)^{-2}(\1-F^\prime)^{-1}\\
&-16{\rm Tr}(\1+F)^{-1}(\1-F^\prime)^{-2}\\
&+16{\rm Tr}(\1+F)^{-1}(\1-F^\prime)^{-1}+4{\rm Tr}(\1+F)^{-2}\\
&+4{\rm Tr}(\1-F^\prime)^{-2}-4{\rm Tr}(\1+F)^{-1}\\
&-4{\rm Tr}(\1-F^\prime)^{-1}+4.
\end{aligned}
\end{equation}

Firstly, we calculate the first term in \eqref{73 TrW2 expanded} by power expansion as
\begin{equation}\label{74sum}
\begin{aligned}
&{\rm Tr}(\1+F)^{-1}(\1-F^\prime)^{-1}(\1+F)^{-1}(\1-F^\prime)^{-1}\\&=\sum_{m,n,p,q=0}^{+\infty}(-1)^{m+q}T_{m,n,p,q},
\end{aligned}
\end{equation}
where $T_{m,n,p,q}$ stands for ${\rm Tr}(F^m{F^\prime}^nF^p{F^\prime}^q)$. Using \eqref{recursion of F} we have recursion relations as
\begin{equation}
\begin{aligned}
&T_{m,n,p,q}=aT_{m,n,p-2,q}+b^2T_{m,n,p-4,q},\\
&T_{m,n,p,q}=a^\prime T_{m,n,p,q-2}+{b^\prime}^2T_{m,n,p,q-4}.
\end{aligned}
\end{equation}
We can calculate all the terms in \eqref{74sum} by solving this second-order recursion problem,
\begin{equation}
\begin{aligned}
T_{m,n,2k,2l}=&T_{m,n,2,2}Q_k{Q^\prime}_l+T_{m,n,0,2}b^2Q_{k-1}{Q^\prime}_l\\
&+T_{m,n,2,0}{b^\prime}^2Q_k{Q^\prime}_{l-1}\\&+T_{m,n,0,0}b^2{b^\prime}^2Q_{k-1}{Q^\prime}_{l-1},\\
T_{m,n,2k-1,2l-1}=&T_{m,n,1,1}Q_k{Q^\prime}_l+T_{m,n,-1,1}b^2Q_{k-1}{Q^\prime}_l\\
&+T_{m,n,1,-1}{b^\prime}^2Q_k{Q^\prime}_{l-1}\\&+T_{m,n,-1,-1}b^2{b^\prime}^2Q_{k-1}{Q^\prime}_{l-1},\\
T_{m,n,2k,2l-1}=&T_{m,n,2,1}Q_k{Q^\prime}_l+T_{m,n,0,1}b^2Q_{k-1}{Q^\prime}_l\\
&+T_{m,n,2,-1}{b^\prime}^2Q_k{Q^\prime}_{l-1}\\&+T_{m,n,0,-1}b^2{b^\prime}^2Q_{k-1}{Q^\prime}_{l-1},\\
T_{m,n,2k-1,2l}=&T_{m,n,1,2}Q_k{Q^\prime}_l+T_{m,n,-1,2}b^2Q_{k-1}{Q^\prime}_l\\
&+T_{m,n,1,0}{b^\prime}^2Q_k{Q^\prime}_{l-1}\\&+T_{m,n,-1,0}b^2{b^\prime}^2Q_{k-1}{Q^\prime}_{l-1}.
\end{aligned}
\end{equation}
Summing over $k$ and $l$, we get
\begin{equation}\label{77 sum over k and l}
\begin{aligned}
&{\rm Tr}(\1+F)^{-1}(\1-F^\prime)^{-1}(\1+F)^{-1}(\1-F^\prime)^{-1}\\
=&\sum_{m,n=0}^{+\infty}(-1)^m\Big(\sum_{k,l=0}^{+\infty}T_{m,n,2k,2l}-\sum_{k,l=1}^{+\infty}T_{m,n,2k-1,2l-1}\\
&+\sum_{k=0,l=1}^{+\infty}T_{m,n,2k,2l-1}-\sum_{k=1,l=0}^{+\infty}T_{m,n,2k-1,2l}\Big)\\
=&QQ^\prime\sum_{m,n=0}^{+\infty}(-1)^m\Big[T_{m,n,2,2}+(1-a)T_{m,n,0,2}\\
&+(1-a^\prime)T_{m,n,2,0}+(1-a)(1-a^\prime)T_{m,n,0,0}-T_{m,n,1,1}\\
&-b^2T_{m,n,-1,1}-{b^\prime}^2T_{m,n,1,-1}-b^2{b^\prime}^2T_{m,n,-1,-1}+T_{m,n,2,1}\\
&+(1-a)T_{m,n,0,1}+{b^\prime}^2T_{m,n,2,-1}+(1-a){b^\prime}^2T_{m,n,0,-1}\\
&-T_{m,n,1,2}-b^2T_{m,n,-1,2}-(1-a^\prime)T_{m,n,1,0}\\
&-(1-a^\prime)b^2T_{m,n,-1,0}\Big].
\end{aligned}
\end{equation}

For a reminder here, since we have defined $T_{m,n}={\rm Tr}(F^m{F^\prime}^n)$ and $T_{m,n,p,q}={\rm Tr}(F^m{F^\prime}^nF^p{F^\prime}^q)$, some terms in the summation of the expression after the last equal sign of \eqref{77 sum over k and l} such as $T_{m,n,0,0}$ can be expressed by $T_{m,n}$ as follows,
\begin{equation}\label{30 T4}
\begin{aligned}
&T_{m,n,0,0}=T_{m,n},\quad T_{m,n,0,2}=T_{m,n+2},\quad T_{m,n,2,0}=T_{m+2,n},\\
&T_{m,n,1,0}=T_{m+1,n},\quad T_{m,n,0,1}=T_{m,n+1},\\
&T_{m,n,-1,0}=T_{m-1,n},\quad T_{m,n,0,-1}=T_{m,n-1}.
\end{aligned}
\end{equation}
Other terms in the summation of the expression after the last equal sign of \eqref{77 sum over k and l} such as $T_{m,n,1,1}$ can also be expressed by $T_{m,n}$ by repeatedly using \eqref{61F*F}. Taking $T_{m,n,1,1}$ as an example, we have
\begin{equation}\label{Tmn11}
\begin{aligned}
T_{m,n,1,1}&={\rm Tr}\big(F^m{F^\prime}^n(\star F^\prime)(\star F)\big)+cT_{m,n}\\
&=bb^\prime T_{m-1.n-1}+cT_{m,n}.
\end{aligned}
\end{equation}
In the first line of \eqref{Tmn11} we have used the fifth relation in \eqref{61F*F}, while in the second line of \eqref{Tmn11} we have used the third and fourth relation in \eqref{61F*F}. The other results of terms needed in the summation of the expression after the last equal sign of \eqref{77 sum over k and l} can be obtained similarly. The results are listed below.
\begin{small}
\begin{equation}\label{32 T4}
\begin{aligned}
&T_{m,n,-1,1}=-\frac{b^\prime}{b}T_{m+1,n-1}+\frac{d}{b}T_{m,n},\\
&T_{m,n,1,-1}=-\frac{b}{b^\prime}T_{m-1,n+1}+\frac{d}{b^\prime}T_{m,n},\\
&T_{m,n,-1,-1}=\frac{1}{bb^\prime}T_{m+1,n+1}-\frac{c}{bb^\prime}T_{m,n},\\
&T_{m,n,2,-1}=-T_{m+2,n-1}+aT_{m,n-1}-\frac{bc}{b^\prime}T_{m-1,n}+\frac{d}{b^\prime}T_{m+1,n},\\
&T_{m,n,-1,2}=-T_{m-1,n+2}+a^\prime T_{m-1,n}-\frac{b^\prime c}{b}T_{m,n-1}+\frac{d}{b}T_{m,n+1},\\
&T_{m,n,2,1}=-T_{m+2,n+1}+aT_{m,n+1}+dbT_{m-1,n}+cT_{m+1,n},\\
&T_{m,n,1,2}=-T_{m+1,n+2}+a^\prime T_{m+1,n}+db^\prime T_{m,n-1}+cT_{m,n+1},\\
&T_{m,n,2,2}=T_{m+2,n+2}-cT_{m+1,n+1}-dbT_{m-1,n+1}\\
&\quad\quad\quad\quad\quad-db^\prime T_{m+1,n-1}+(c^2+d^2)T_{m,n}+cbb^\prime T_{m-1,n-1}.
\end{aligned}
\end{equation}
\end{small}
Also, we choose $p$ and $q$ starting from ($-1$) because of simplicity, and the final results still holds even if $b$ or $b^\prime$ is zero. With all terms needed in the summation of the expression after the last equal sign of \eqref{77 sum over k and l} expressed in terms of $T_{m,n}$ in \eqref{30 T4}, \eqref{Tmn11} and \eqref{32 T4}, as well as $T_{mn}$ already given in \eqref{67 Tmn}, we can substitute these results into \eqref{77 sum over k and l} to get
\begin{small}
\begin{equation}\label{81 sum over m and n}
\begin{aligned}
&{\rm Tr}(\1+F)^{-1}(\1-F^\prime)^{-1}(\1+F)^{-1}(\1-F^\prime)^{-1}\\
=&Q^2{Q^\prime}^2\big[(2-a)(2-a^\prime)+c^2+d^2-2c(1-bb^\prime)-2d(b+b^\prime)\big]^2\\
&+2QQ^\prime\big[2bb^\prime-(2-a)-(2-a^\prime)-c^2+4c-2\big].
\end{aligned}
\end{equation}
\end{small}

We are now calculating other terms needed in \eqref{73 TrW2 expanded}. Since some of them are already given in \eqref{equationNo70} and \eqref{equationNo71}, we are calculating those not given in \eqref{equationNo70} and \eqref{equationNo71}, by power expansion as
\begin{equation}\label{39TTTT}
\begin{aligned}
&{\rm Tr}(1+F)^{-2}(1-F^\prime)^{-1}=\sum_{m,n=0}^{+\infty}(-1)^m(m+1)T_{m,n},\\
&{\rm Tr}(1+F)^{-1}(1-F^\prime)^{-2}=\sum_{m,n=0}^{+\infty}(-1)^m(n+1)T_{m,n},\\
&{\rm Tr}(1+F)^{-2}=\sum_{m=0}^{+\infty}(-1)^m(m+1)T_{m,0},\\
&{\rm Tr}(1-F^\prime)^{-2}=\sum_{n=0}^{+\infty}(n+1)T_{0,n}.
\end{aligned}
\end{equation} 
In order to sum over $m$ and $n$ in \eqref{39TTTT}, we need
\begin{equation}
\sum_{i=0}^{+\infty}iQ_i=\frac{1}{\alpha-\beta}\Big(\frac{\alpha}{(1-\alpha)^2}-\frac{\beta}{(1-\beta)^2}\Big)=Q^2(1+b^2),
\end{equation}
similar for the summation $\sum_{i=0}^{+\infty}i{Q^\prime}_i$. Using results of $T_{m,n}$ given in \eqref{67 Tmn}, expressions in \eqref{39TTTT} can be expressed in terms of the six Lorentz invariants as

\begin{equation}\label{36TTTTdetailed}
\begin{aligned}
&{\rm Tr}(1+F)^{-2}(1-F^\prime)^{-1}\\
&=QQ^\prime\big[(2-a)(2-a^\prime)+c^2+d^2\big]\\
&\quad-2QQ^\prime\big[(4-a)(2-a^\prime)+c^2+d^2-2c-2db^\prime\big]\\
&\quad+2Q^2Q^\prime(2-a)\big[(2-a)(2-a^\prime)+c^2+d^2\\
&\quad\quad\quad\quad\quad\quad\quad\quad-2c(1-bb^\prime)-2d(b+b^\prime)\big],\\
&{\rm Tr}(1+F)^{-1}(1-F^\prime)^{-2}\\
&=QQ^\prime\big[(2-a)(2-a^\prime)+c^2+d^2\big]\\
&\quad-2QQ^\prime\big[(2-a)(4-a^\prime)+c^2+d^2-2c-2db\big]\\
&\quad+2Q{Q^\prime}^2(2-a^\prime)\big[(2-a)(2-a^\prime)+c^2+d^2\\
&\quad\quad\quad\quad\quad\quad\quad\quad-2c(1-bb^\prime)-2d(b+b^\prime)\big],\\
&{\rm Tr}(1+F)^{-2}=4Q^2(2-a)^2-2Q(6-a),\\
&{\rm Tr}(1-F^\prime)^{-2}=4{Q^\prime}^2(2-a^\prime)^2-2Q^\prime(6-a^\prime).
\end{aligned}
\end{equation}
Substitute results in \eqref{equationNo70}, \eqref{equationNo71},\eqref{81 sum over m and n} and \eqref{36TTTTdetailed} into \eqref{73 TrW2 expanded}, we can derive the result of ${\rm Tr}(W^2)$ as

\begin{small}
\begin{equation}
\begin{aligned}
&{\rm Tr}(W^2)=\\
&\quad16Q^2{Q^\prime}^2\big[(2-a)(2-a^\prime)+c^2+d^2\\
&\quad\quad\quad\quad\quad\quad-2c(1-bb^\prime)-2d(b+b^\prime)\big]^2\\
&\quad+32QQ^\prime\big[2bb^\prime-(2-a)-(2-a^\prime)-c^2+4c-2\big]\\
&\quad-32Q^2Q^\prime(2-a)\big[(2-a)(2-a^\prime)+c^2+d^2-2c(1-bb^\prime)\\
&\quad\quad\quad\quad\quad\quad\quad\quad\quad-2d(b+b^\prime)\big]\\
&\quad-32Q{Q^\prime}^2(2-a^\prime)\big[(2-a)(2-a^\prime)+c^2+d^2-2c(1-bb^\prime)\\
&\quad\quad\quad\quad\quad\quad\quad\quad\quad-2d(b+b^\prime)\big]\\
&\quad-16QQ^\prime\big[(2-a)(2-a^\prime)+c^2+d^2+2c(1-bb^\prime)+2d(b+b^\prime)\big]\\
&\quad-32QQ^\prime\big[-(4-a)(2-a^\prime)-(2-a)(4-a^\prime)\\
&\quad\quad\quad\quad\quad\quad-2c^2-2d^2+4c+2d(b+b^\prime)\big]\\
&\quad+16Q^2(2-a)^2-16Q(4-a)+16{Q^\prime}^2(2-a^\prime)^2\\
&\quad-16Q^\prime(4-a^\prime)+4.
\end{aligned}
\end{equation}
\end{small}

Taking the square of the expression after the first equal sign in \eqref{72 TrW}, where we gave the result of Tr$W$, we get
\begin{small}
\begin{equation}
\begin{aligned}
&({\rm Tr}W)^2\\
&\quad=16Q^2{Q^\prime}^2\big[(2-a)(2-a^\prime)+c^2+d^2-2c(1-bb^\prime)\\
&\quad\quad\quad\quad\quad\quad-2d(b+b^\prime)\big]^2\\
&\quad-32Q^2Q^\prime(2-a)\big[(2-a)(2-a^\prime)+c^2+d^2-2c(1-bb^\prime)\\
&\quad\quad\quad\quad\quad\quad\quad\quad\quad-2d(b+b^\prime)\big]\\
&\quad-32Q{Q^\prime}^2(2-a^\prime)\big[(2-a)(2-a^\prime)+c^2+d^2-2c(1-bb^\prime)\\&
\quad\quad\quad\quad\quad\quad\quad\quad\quad-2d(b+b^\prime)\big]\\
&\quad+32QQ^\prime\big[(2-a)(2-a^\prime)+c^2+d^2-2c(1-bb^\prime)-2d(b+b^\prime)\big]\\
&\quad+16Q^2(2-a)^2+16{Q^\prime}^2(2-a^\prime)^2+32Q{Q^\prime}(2-a)(2-a^\prime)\\
&\quad-32Q(2-a)-32Q^\prime(2-a^\prime)+16.
\end{aligned}
\end{equation}
\end{small}

With lots of complicated terms canceled, the final result is quite simple,
\begin{equation}\label{TrW2-TrW2}
\begin{aligned}
&({\rm Tr}W)^2-{\rm Tr}(W^2)\\
=&16QQ^\prime\big[aa^\prime+c^2-d^2-2c(1-bb^\prime)+2d(b+b^\prime-4bb^\prime)\big]\\
&+16Qa+16Q^\prime a^\prime+12\\
=&16QQ^\prime\big[(1-b^2)(1-{b^\prime}^2)+c^2-d^2-2c(1-bb^\prime)\\&+2d(b+b^\prime-4bb^\prime)\big]-4.
\end{aligned}
\end{equation}
Also, in the derivation we have used $(1-a-b^2)Q=(1-a^\prime-{b^\prime}^2)Q^\prime=1$. Substituting \eqref{72 TrW} and \eqref{TrW2-TrW2} to \eqref{52c2c2} and \eqref{53s2s2}, we get
\begin{equation}
\begin{aligned}
&{\rm cos}^2\pi \nu_0{\rm cos}^2\pi \nu_1=QQ^\prime(1-c-bb^\prime)^2,\\
&{\rm sin}^2\pi \nu_0{\rm sin}^2\pi \nu_1=-QQ^\prime(\Delta b)^2.
\end{aligned}
\end{equation}
\\[25pt]
\section{Eigenvalue structure and diagonalization of $SO(p,1)$ matrices}\label{Appendix B}
\indent In this paper we have claimed that the unitary matrix $W$ has eigenvalues take the form $e^{2\pi i\nu_0}$ and $e^{2\pi i \nu_1}$, where $\nu_0$ and $\nu_1$ are one pure imaginary (or zero) and the other real. In fact, we have similar properties for a unitary matrix $W$ whose dimensionality is arbitrary, in Minkowski metric. In this appendix, the dimensionality of $W$ is no longer 4 but $p+1$, where $p$ is an arbitrary positive integer. If $p$ is odd, the eigenvalues of $W$ will take the form as $e^{\pm2\pi i\nu_0}$, $e^{\pm2\pi i \nu_1}$, $\cdots$, $e^{\pm2\pi i \nu_{\frac{p+1}{2}}}$. Among $\nu_0\sim \nu_{\frac{p+1}{2}}$, only one of them is pure imaginary (or zero) and the others are real. If $p$ is even, the eigenvalues of $W$ will take the form as 1,$e^{\pm2\pi i\nu_0}$, $e^{\pm2\pi i \nu_1}$, $\cdots$, $e^{\pm2\pi \nu_{\frac{p}{2}}}$. Among $\nu_0\sim \nu_{\frac{p}{2}}$, either all of them are real, or one of them is pure imaginary (or zero) and the others are real. We can prove this by the quasi-diagonalization of $W$. 

Since $W\in SO(p,1)$, it can be linearized as $W=e^{K}$, where $K$ is an antisymmetric real matrix that satisfies $K^{\rm{T}}=-K$. We now try to quasi-diagonalize $K$ via a series of Lorentz transforms to render it to a standard quasi-diagonalized form. For odd $p$, the standard form is
\begin{equation}\label{quasi-diagonalization of K}
K_0=\Lambda^{-1}K\Lambda=
\begin{pmatrix}
0&k_0&&&&&\\
k_0&0&&&&&\\
&&0&-k_1&&&\\
&&k_1&0&&&\\
&&&&\ddots&&\\
&&&&&0&-k_{\frac{p+1}{2}}\\
&&&&&k_{\frac{p+1}{2}}&0
\end{pmatrix},
\end{equation}
where $\Lambda\in SO(p,1)$. For even $p$, the standard form is either
\begin{equation}\label{quasi-diagonalization of K,1}
K_0=\Lambda^{-1}K\Lambda=
\begin{pmatrix}
0&k_0&&&&&&\\
k_0&0&&&&&&\\
&&0&-k_1&&&&\\
&&k_1&0&&&&\\
&&&&\ddots&&&\\
&&&&&0&-k_{\frac{p}{2}}&\\
&&&&&k_{\frac{p}{2}}&0&\\
&&&&&&&0
\end{pmatrix}.
\end{equation}
or
\begin{equation}\label{quasi-diagonalization of K,2}
K_0=\Lambda^{-1}K\Lambda=
\begin{pmatrix}
0&&&&&&&\\
&0&-k_0&&&&&\\
&k_0&0&&&&&\\
&&&0&-k_1&&&\\
&&&k_1&0&&&\\
&&&&&\ddots&&\\
&&&&&&0&-k_{\frac{p}{2}}\\
&&&&&&k_{\frac{p}{2}}&0
\end{pmatrix}.
\end{equation}
We are going to realize this in following steps.\\\indent
In step 1, our purpose is to render 
\begin{equation}
{(H_1)_0}^i=0,
\end{equation}
where $i,j,\cdots=1,2,3,\cdots$ and
\begin{equation}
{H_\alpha}^\beta\overset{\rm def}{=}{K_\alpha}^\gamma{K_\gamma}^\beta,
\end{equation}
where $\alpha,\beta,\cdots=0,1,2,3,\cdots$, by performing a Lorentz boost as
\begin{equation}
K_1=(\Lambda_1)^{-1}K\Lambda_1,
\end{equation}
where
\begin{equation}\label{65LL}
{(\Lambda_1)_0}^0={\rm cosh}t,
\end{equation}
\begin{equation}\label{66LL}
{(\Lambda_1)_0}^i={(\Lambda_1)_i}^0=n^i{\rm sinh}t,
\end{equation}
\begin{equation}\label{67LL}
{(\Lambda_1)_i}^j=({\rm cosh}t-1)n_in^j+{\delta_i}^j,
\end{equation}
where $n_in^i=1$. If ${H_0}^i=0$ is already satisfied, we just skip this step (that is $\Lambda_1=\1$). After doing so, we require
\begin{equation}
\begin{aligned}\label{H0i=0}
0=&{(H_1)_0}^i={({\Lambda_1}^{-1})_0}^\alpha{H_\alpha}^\beta{(\Lambda_1)_\beta}^i\\
=&\frac{1}{2}{\rm sinh}2t{H_0}^0n^i+({\rm cosh}2t-{\rm cosh}t){H_0}^jn_jn^i+{\rm cosh}t{H_0}^i\\
&-{\rm sinh}t({\rm cosh}t-1)(n^j{H_j}^kn_k)n^i-{\rm sinh}t{H_j}^in^j.
\end{aligned}
\end{equation}
This equation is solvable since it has $p$ equations and $p$ independent unknowns. Later in this appendix, we will solve this equation while $p=3$ as an example.

After step 1 we have made ${(H_1)_0}^i=0$. Then, in step 2, we perform a pure spatial rotation as follows
\begin{equation}
\Lambda_2=
\begin{pmatrix}
1&\\
&{R_i}^j
\end{pmatrix},
\end{equation}
and
\begin{equation}
K_2=(\Lambda_2)^{-1}K_1\Lambda_2.
\end{equation}
Under such a transformation, ${K_0}^i$ and ${H_0}^i$ are rotated as $p$-dimensional vectors. There are two sub cases.\\\indent
If ${(K_1)_0}^i{(K_1)_0}^i\neq0$, we can rotate ${(K_1)_0}^i$ to direction ``1", which means ${(K_2)_0}^i=0$ for $i\neq1$ and ${(K_2)_0}^1\neq0$. Since ${(H_1)_0}^i=0$, after the rotation we also have ${(H_2)_0}^i=0$, which means 
\begin{equation}
0={(H_2)_0}^i={(K_2)_0}^\alpha{(K_2)_\alpha}^i={(K_2)_0}^1{(K_2)_1}^i.
\end{equation}
Since ${K_0}^1\neq0$, we have ${(K_2)_1}^i=0$. Thus, $K_2$ takes the form
\begin{equation}
K_2=
\begin{pmatrix}
0&{(K_2)_0}^1&\\
{(K_2)_0}^1&0&\\
&&{(K_2)_a}^b
\end{pmatrix},
\end{equation}
where $a,b,\cdots=2,3,4,\cdots$ \\
\indent
If ${(K_1)_0}^i{(K_1)_0}^i=0$, which means $K_1$ takes the form
\begin{equation}
K_1=
\begin{pmatrix}
0&\\
&{(K_1)_i}^j
\end{pmatrix}.
\end{equation}

If $p$ is even, we do nothing in step 2. That is
\begin{equation}
K_2=K_1=
\begin{pmatrix}
0&\\
&{(K_2)_i}^j
\end{pmatrix}.
\end{equation}

If $p$ is odd, since any odd-dimensional antisymmetric matrix has at least one null vector, consider the null vector $\zeta^i$ satisfying $(\zeta_1)^i{(K_1)_i}^j=0$, we can rotate it to direction ``1", that is to render $(\zeta_2)^i=0$ for $i\neq1$ and $(\zeta_2)^1\neq0$ , which means ${(K_2)_0}^i=0$. Thus, $K_2$ takes the form 
\begin{equation}
K_0=
\begin{pmatrix}
0&0&\\
0&0&\\
&&{(K_2)_a}^b\\
\end{pmatrix}.
\end{equation}

Following step 1 and 2, we have partially quasi-diagonalized $K$, with a $p$-dimensional antisymmetric matrix ${(K_2)_i}^j$ or $(p-1)$-dimensional antisymmetric matrix ${(K_2)_a}^b$ remaining to be quasi-diagonalized. We can repeat similar procedure in step 1 and 2 again and again in the $p$-dimensional or $(p-1)$-dimensional subspace to get the standard quasi-diagonalized form \eqref{quasi-diagonalization of K}, \eqref{quasi-diagonalization of K,1} or \eqref{quasi-diagonalization of K,2}. Notice that in the next step, namely, step 3, which is similar to step 1, the remaining ${(K_2)_i}^j$ or ${(K_2)_a}^b$ is antisymmetric in Euclidean metric, one should modify the Lorentz boost in step 1, introduced in \eqref{65LL},\eqref{66LL} and \eqref{67LL}, to a rotation as
\begin{equation}
{(\Lambda_3)_1}^1={\rm cos}\theta,
\end{equation}
\begin{equation}
{(\Lambda_3)_1}^a=-{(\Lambda_3)_a}^0=n^a{\rm sin}\theta,
\end{equation}
\begin{equation}
{(\Lambda_3)_a}^b=({\rm cos}\theta-1)n_an^b+{\delta_a}^b,
\end{equation}
for quasi-diagonalization of ${(K_2)_i}^j$ or
\begin{equation}
{(\Lambda_3)_2}^2={\rm cos}\theta,
\end{equation}
\begin{equation}
{(\Lambda_3)_2}^A=-{(\Lambda_3)_A}^0=n^A{\rm sin}\theta,
\end{equation}
\begin{equation}
{(\Lambda_3)_A}^B=({\rm cos}\theta-1)n_An^B+{\delta_A}^B,
\end{equation}
for quasi-diagonalization of ${(K_2)_a}^b$, where $A,B\cdots=3,4,5,\cdots$. 

Following the procedures above, we quasi-diagonalized $K$. If $p$ is odd, we have
\begin{equation}
K_0=\Lambda^{-1}K\Lambda=
\begin{pmatrix}
0&k_0&&&&&\\
k_0&0&&&&&\\
&&0&-k_1&&&\\
&&k_1&0&&&\\
&&&&\ddots&&\\
&&&&&0&-k_{\frac{p+1}{2}}\\
&&&&&k_{\frac{p+1}{2}}&0
\end{pmatrix}.
\end{equation}
In this case, the eigenvalues of $K$ are $\pm k_0$, $\pm ik_1$, $\cdots$, $\pm ik_{\frac{p+1}{2}}$, with $k_\alpha$ all being real. Then, the eigenvalues of $W$ are $e^{\pm2\pi i \nu_0}$ and $e^{\pm2\pi i \nu_1}$, $\cdots$, $e^{\pm2\pi i \nu_\frac{p+1}{2}}$, with $\nu_0=\frac{ik_0}{2\pi}$ and $\nu_i=\frac{k_i}{2\pi}$.

If $p$ is even, there are two sub cases. In sub case 1, we have
\begin{equation}
K_0=\Lambda^{-1}K\Lambda=
\begin{pmatrix}
0&k_0&&&&&&\\
k_0&0&&&&&&\\
&&0&-k_1&&&&\\
&&k_1&0&&&&\\
&&&&\ddots&&&\\
&&&&&0&-k_{\frac{p}{2}}&\\
&&&&&k_{\frac{p}{2}}&0&\\
&&&&&&&0
\end{pmatrix}.
\end{equation}
In this sub case, the eigenvalues of $K$ are $\pm k_0$, $\pm ik_1$, $\cdots$, $\pm ik_{\frac{p+1}{2}}$, and 0, with $k_\alpha$ all being real. Then, the eigenvalues of $W$ are $e^{\pm2\pi i \nu_0}$ and $e^{\pm2\pi i \nu_1}$, $\cdots$, $e^{\pm2\pi i \nu_\frac{p+1}{2}}$,and 1, with $\nu_0=\frac{ik_0}{2\pi}$ and $\nu_i=\frac{k_i}{2\pi}$.

In sub case 2, we have
\begin{equation}
K_0=\Lambda^{-1}K\Lambda=
\begin{pmatrix}
0&&&&&&&\\
&0&-k_0&&&&&\\
&k_0&0&&&&&\\
&&&0&-k_1&&&\\
&&&k_1&0&&&\\
&&&&&\ddots&&\\
&&&&&&0&-k_{\frac{p}{2}}\\
&&&&&&k_{\frac{p}{2}}&0
\end{pmatrix}.
\end{equation}
In this sub case, the eigenvalues of $K$ are 0, $\pm i k_0$, $\pm ik_1$, $\cdots$, $\pm ik_{\frac{p+1}{2}}$, with $k_\alpha$ all being real. Then, the eigenvalues of $W$ are 1, $e^{\pm2\pi i \nu_0}$ and $e^{\pm2\pi i \nu_1}$, $\cdots$, $e^{\pm2\pi i \nu_\frac{p+1}{2}}$, with $\nu_\alpha=\frac{k_\alpha}{2\pi}$.

Let us take $p=3$ case as an example. In such a case, in order to derive $\Lambda_1$, we need to solve \eqref{H0i=0}. The solution for $n_i$ is
\begin{equation}
n^i=\frac{{H_0}^i}{\sqrt{{H_0}^j{H_0}_j}}.
\end{equation}
When $p=3$, this solution satisfies
\begin{equation}
{H_j}^in^j=(n^j{H_j}^kn_k)n^i.
\end{equation}
Substituting these to \eqref{H0i=0} gives the solution of $t$ as
\begin{equation}\label{sov of t}
{\rm tanh}2t=-\frac{2{H_0}^in_i}{{H_0}^0-(n^j{H_j}^kn_k)}.
\end{equation}
Such a $t$ always exists. Take $U^\alpha={K_0}^\alpha$ and $V^\alpha=n^i{K_i}^\alpha$, since $U^0=V^0=0$, we have $U^\alpha U_\alpha+V^\alpha V_\alpha \ge 2|U^\alpha V_\alpha|$. Thus, 
\begin{equation}
[{H_0}^0-(n^j{H_j}^kn_k)]\ge2|{H_0}^in_i|.
\end{equation}
Therefore, we can always find such a $t$ that satisfies \eqref{sov of t}. Thus, we have set ${(H_1)_0}^i$=0 via $\Lambda_1$. Then, we follow step 2 to get
\begin{equation}
K_0=\Lambda^{-1}K\Lambda=
\begin{pmatrix}
0&k_0&&\\
k_0&0&&\\
&&0&-k_1\\
&&k_1&0
\end{pmatrix},
\end{equation}
with $\Lambda=\Lambda_1\Lambda_2$. The eigenvalues of $K$ is clearly $\pm k_0$ and $\pm ik_1$. Then, the eigenvalues of $W$ is $e^{\pm2\pi i \nu_0}$ and $e^{\pm2\pi i \nu_1}$, with $\nu_0=\frac{ik_0}{2\pi}$ and $\nu_1=\frac{k_1}{2\pi}$. 

\nocite{*}
\bibliography{bibtex}
\bibliographystyle{unsrt}

\end{document}